\documentclass[final,12pt]{article}
\usepackage[utf8]{inputenc}
\usepackage{anysize}
\marginsize{2cm}{2cm}{2cm}{2cm} 

\usepackage{fancyhdr} 
\usepackage{lipsum} 
\usepackage{color} 
\usepackage{bm} 
\usepackage{amsfonts,amsthm,amsmath,amssymb} 
\usepackage{latexsym,graphicx,epstopdf} 
\usepackage{etoolbox} 
\usepackage{hyperref}

\newcommand{\n}{{\bf n}}

\renewcommand{\u}{{\bf u}}

\newcommand{\ff}{{\bf F}}

\title{The energy of muscle contraction. II. Transverse compression and work.\thanks{This work was partially supported by the Natural Sciences and Engineering Research Council of Canada and the Comisi\'on Nacional de Investigatici\'on Cient\'ifica y Tecnol\'ogica of Chile.}}

\author{David S. Ryan\thanks{Department of Biomedical Physiology and Kinesiology, Simon Fraser University, Burnaby, BC, Canada}\quad Sebasti\'an Dom\'inguez{\,\,\,\thanks{Department of Mathematics, Simon Fraser University, Burnaby, BC, Canada}}{\,\,\,\thanks{Corresponding author: 
\href{mailto:domingue@sfu.ca}{domingue@sfu.ca}}}\quad Stephanie A. Ross\footnotemark[2]\\ 
Nilima Nigam\footnotemark[3]\quad James M. Wakeling\footnotemark[2]
}

\begin{document}

\maketitle

\begin{abstract}
In this study we reproduced this compression-induced reduction in muscle force through the use of a
three-dimensional finite element model of contracting muscle. The model used the principle of
minimum total energy and allowed for the redistribution of energy through different strain energy-
densities; this allowed us to determine the importance of the strain energy-densities to the transverse
forces developed by the muscle. Furthermore, we were able to study how external work done on the
muscle by transverse compression affects the internal work and strain-energy distribution of the
muscle. We ran a series of in silica experiments on muscle blocks varying in initial pennation angle,
muscle length, and compressive load. As muscle contracts it maintains a near constant volume. As
such, any changes in muscle length are balanced by deformations in the transverse directions such as
muscle thickness or muscle width. Muscle develops transverse forces as it expands. In many
situations external forces work to counteract these transverse forces. Muscle responds to external
transverse compression while both passive and active. Transverse compression leads to a reduction in
muscle thickness and pennation angle when the muscle is passive, and a reduction to the longitudinal
force in its line-of-action when the muscle is active. Greater transverse compression leads to greater
force reduction. The muscle blocks used in our simulations decreased in thickness and pennation
angle when passively compressed, and pushed back on the compression when they were activated.
We show how the longitudinal force from the muscle reduces with increased compressive load and
that this reduction is dependent on the pennation angle and muscle length. The compression-induced
reductions in the longitudinal muscle force were largely due to the volumetric strain-energy density,
which is function of the bulk modulus of the muscle tissue and the dilation of the tissue.

\end{abstract}


\vspace{.25cm}


\section{Introduction}\label{section:intro}
Muscles change in length and develop longitudinal force when they contract, and these result in
internal work being done within the muscle. Muscles additionally expand and develop forces in
transverse directions, again resulting from internal work. However, the transverse action of muscle is
rarely studied. In this paper we show how longitudinal and transverse forces and deformations of the muscle are coupled via the internal energy of the muscle, and in particular through the redistribution
of energy across different forms of strain-energy potentials.

Shape changes and muscle forces occur in all three dimensions when muscles contract. As a
muscle shortens, it must increase in girth or cross-sectional area in order to maintain its volume
(Zuurbier and Huijing, 1993; Böl et al., 2013; Randhawa and Wakeling, 2015). Transverse
expansions of contracting muscle have been reported in both animal (Brainerd and Azizi, 2005; Azizi
et al., 2008) and human studies (Maganaris et al. 1998; Randhawa et al., 2013; Dick and Wakeling,
2017), and transverse forces generated internally in the muscle can ‘lift’ weights during contraction
(Siebert et al. 2014). Conversely, transverse loads that compress the muscle in its cross-section
should be transferred to forces and changes in length along the line-of-action of the muscle.

The force that a muscle develops in its line-of-action depends on pressure and external loads
applied in the transverse direction. Various researchers have used models of fibre-wound helical
tubes (mimicking the endomysium and perimysium of the extracellular matrix) to explain the transfer
of radial to longitudinal forces and deformations in the muscle (Azizi et al. 2017; Sleboda et al. 2017,
2019). Loading the extracellular matrix by increasing the volume of the semimembranosus muscle of
the bullfrog, using osmotic pressure, increases the passive force in the line-of-action of the muscle
(Sleboda et al. 2017; 2019). Limiting radial expansion of muscle by more circumferentially oriented
fibres in the helix reduces the extent to which muscle can shorten, and placing a stiff tube around
contracting frog plantaris muscle reduces both how much the muscle shortens and work done in the
line-of-action (Azizi et al. 2017).

Transverse external loads act to compress passive muscle as they do mechanical work on the
tissue. Compression of passive muscle has been described for isolated medial gastrocnemius in rats
(Siebert et al. 2014, 2016) and for gluteus maximus (Linder-Ganz et al. 2007) and medial
gastrocnemius (Stutzig et al. 2019; Ryan et al. 2019) muscles in humans. Muscles bulge to resist the
transverse loads when they activate and work is generated from forces that develop in the transverse
direction. The work generated from these transverse forces can be thought of as ‘lifting work’,
especially if it is working against gravity (see the Methods for the formal definition). The muscle
volume-specific energy involved in this ‘lifting work’ from the medial gastrocnemius has been
approximately 1.1-1.2 x 10 3 J m -3 (Siebert et al. 2014) in the rat, and 1.1 x 10 3 J m -3 in humans
(Stutzig et al. 2019): it should be noted that in these experiments the plungers that applied the
transverse load covered only about 20
is about 2 orders of magnitude less than the work that could be done by the longitudinal muscle force
(Weis-Fogh and Alexander, 1977). This force in the line-of-action during muscle contraction is
reduced when the muscle does work to resist the transverse loads (Siebert et al. 2014; 2018; Stutzig
et al. 2019; Ryan et al. 2019), and the extent of this force reduction depends on the transverse force
rather than the external load applied to the muscle (Siebert et al. 2016). Siebert and colleagues (2012)
explained transverse muscle forces and bulging from previous data using a hydraulically driven
model that transfers load between the transverse to longitudinal directions. They used an ellipsoidal
geometry with constraints that governed anisotropy in the deformations: their model indicated that
anisotropy in the connective tissue was important for the transfer of loads between the transverse and
longitudinal directions.

Muscles are additionally packaged together in anatomical compartments, and they squeeze on
each other as they bulge during contraction. This caused a decrease in the force from the combined
quadriceps in the rabbit when compared to the sum of the individual muscle forces if they were
stimulated separately (de Brito Fontana et al. 2018, 2020), although the reasons for this were not clear. Not all muscles increase in thickness during isometric contractions, and thus we should not
expect that every muscle will squeeze into neighbouring muscles when they activate. Muscles with
lower pennation angles ($<15$ degrees) tend to bulge, whereas more pennate muscle may thin as they activate
(Randhawa et al. 2013; Wakeling et al. 2020).

The changes in muscle force with external loads are length-dependent. For example, the force
in bullfrog semimembranosus was decreased at short lengths and increased at long lengths, when
compared to the resting length, when a pressure cuff was applied around the muscle to apply
transverse force (Sleboda et al. 2019). Sleboda and colleagues (2019) explained these findings using
a helically wound model in which the muscle acts to return to a length at which the angle of the
helical fibres returns to their ideal pitch of $55$ degrees (Wainwright et al. 1976), and this pitch was assumed
to occur at the muscle’s resting length. In contrast, greater reductions in muscle force have been
detected in human plantarflexor muscles when they are compressed while at longer lengths (knee
extended; Siebert et al. 2018) than at a shorter length (knee flexed: Ryan and Siebert observations).

When muscles activate, they increase in their pennation angle both for shortening and for
isometric contractions. Internal deformations additionally occur within muscle when it is
compressed: external transverse loads cause a reduction in the mean fibre pennation angle (Wakeling
et al. 2013), and a reduction to the extent the pennation angle increases when the muscle contracts
(Ryan et al. 2019). We previously showed that the redistribution of strain-energy potentials within
contracting muscle changes with the pennation angle (Wakeling et al. 2020), and it is likely that work
done on and by the muscle generated by forces in the transverse direction would also affect the
strain-energy potentials within the muscle. Thus, we would expect that the external compression
affects the strain-energy potentials within the muscle, that in turn could explain the changes in force
in the line-of-action. The redistribution between the forms of energy also depends on the muscle
length (Wakeling et al. 2020), and this may drive an interaction between the muscle length and the
force reduction that occurs with external load, however, this has not yet been examined. It is also
likely that the strain-energy potentials and the transfer of external loads on the muscles will depend
on the direction of the external load relative to the fibre pennation (for example, is the muscle
compressed from its top or from its side within the transverse plane).

These recent studies have shown that muscle force changes when external transverse loads are
applied to the muscle, and that this effect is length dependent. Theories have discussed how
transverse forces are transferred to longitudinal forces through the properties of the connective tissue
in the muscle (Smith et al., 2011; Sleboda et al. 2019). However, these studies have not explained
how the defomation of muscle tissues due to external compression loads is affected by the internal
geometry or by the direction of the applied load relative to the muscle fibres. Our previous
description of muscle (Wakeling et al. 2020), that quantifies how strain-energy potentials in the
contractile elements are redistributed throughout the muscle volume and the contractile force is
redirected across the muscle tissue, is well suited to understand these mechanics of muscle
compression. The purpose of this study was to identify whether the altered muscle forces that occur
with compression can be explained in terms of the strain-energy potentials within the muscle, and in
particular to account for the role of muscle length, pennation angle and the direction of the external
load on the changes to muscle force.


\section{Methods}\label{section:methods}
In this paper we present simulations to compare the changes in the internal energy and internal
pressure of muscle tissue during an external compression. We modelled the muscle as a three-
dimensional and nearly incompressible fibre-reinforced composite biomaterial. The presence of 1D
fibres through the base-material, representing contractile elements, results in an overall anisotropic
response of the muscle tissue. The formulation of our model used in all simulations is based on the
balance of strain-energy potentials as presented in Wakeling et al. 2020, and was solved using the
finite element method (FEM). The main change is that here we study the influence of external
loading on muscle force output.

The internal pressure is related to the muscle volume and the volumetric strain energy-density $\Psi_{vol}$ in the muscle:
\begin{align}
 \Psi_{vol}(\u,p,J) := \frac{\kappa}{4}\big(J^2-2\log(J) - 1\big) + p\big(J-I_3(\ff)\big),
\end{align}
and can be calculated from the first variation in the volumetric strain energy-density with respect to $J$:
\begin{align*}
 p - \frac{\kappa}{2}\big(J-1/J\big) = 0,
\end{align*}
where $\kappa$ is the bulk modulus of the tissue, $J$ the dilation, and $I_3(\ff)$ the third invariant of the
deformation tensor $\ff$ (Wakeling et al. 2020). Note that we use strain-energy potentials to compare between blocks for the most of the discussion because the muscle blocks in this study all had the same initial volume. Muscle can show small changes in volume when it contracts (Neering et al., 1991; Smith et al., 2011; Bolsterlee et al., 2017), and so we have modelled the muscle as a nearly incompressible tissue (Wakeling et al. 2020).

\subsection{Muscle geometries and simulations}
We constructed a series of blocks of parallel-fibred and unipennate muscle with cuboid
geometries (30x10x10 mm) and no aponeurosis. The origin of the coordinate system was centred
within the blocks for their initial configuration $V_0$. The muscle blocks had faces in the positive and
negative x, y, and z sides. We defined the length of the blocks as the distance between the positive
and negative x-faces in the x-direction, and the thickness as the distance between the positive and
negative z-faces in the z-direction. The muscle fibres were parallel to each other and the xz plane in
$V_0$, but oriented at an initial pennation angle $\beta_0$ (0-40 degrees) away from the x-direction. We set the initial
length of the fibres to their optimal length ($\lambda_{iso} = 1$), and the normalized muscle length $\hat{l}$ to 1 for the
undeformed blocks in their initial configuration $V_0$. The material properties for the muscle tissue were
the same as for our previous study (Wakeling et al. 2020), and we continued to use a scaling factor
$s_base$ of 1.5 for the base-material stiffness.

In silico compression tests were conducted in a series of stages as shown in Fig. 1. The
different steps in our compression tests can be listed as follows:
\begin{itemize}
 \item[(A)] We initially fixed the -x face in all directions, fixed the -z face in the z axis only, and we applied a traction to the +x face to either stretch or shorten the passive muscle. This traction was applied in the direction normal to the +x face in the initial configuration $V_0$.
 \item[(B)] Next, we changed the constraints on the -x face to fix it only in the x-direction, fixed the -z face in all directions and we compressed the passive muscle by applying a transverse load (traction) of 0, 5, 15 or 30 kPa, consistent with previous experimental loads (Ryan et al., 2019; Stutzig et al., 2019). This traction was either on the +z face for ‘top’ loading, or the
traction was from both y-faces for ‘side’ loading.
 \item[(C)] Finally, we fixed both x-faces in the x-direction, maintained the z-constraint on the -z face, and maintained the transverse traction to compress the muscle. During this stage we ramped the activation $\hat{a}$ from 0 to 100\% over a series of 10 time-steps.
\end{itemize}

\subsection{Mechanical and lifting work}
The work done by the muscle tissues during deformation is defined in terms of the force
developed by the muscle, denoted by $F$, at a given point and the displacement $\u$ at the same point. The total work is then defined as
\begin{align*}
 W_{int} := \int_V F\cdot\u\,dV,
\end{align*}
where $V$ is the current configuration of the muscle tissue, and the dot between the vectors denote the dot product. The external work done by the prescribed compression loads on parts of the surface of the muscle geometries, which we denote by $S$, are computed as
\begin{align*}
 W_{ext} := \int_S p_0\hat\n\cdot\u\,dS,
\end{align*}
where $p_0\hat\n$ is the transverse force of the external compression and $\hat\n$ is the normal unit vector on the surface $S$. During activation of the muscle fibres the internal force may be greater than the force on the system from external loads on the surface $S$. In such cases, one can see that the muscle surface pushes back as the external compression load can no longer compress the tissues. The non-zero force that pushes back on the surface of the muscle, denoted by $F_{lift}$, defines a non-zero work which is done by the tissues. We refer to this work as the ‘lifting work’ of the muscle. The ‘lifting work’ is defined as
\begin{align*}
 W_{lift} := \int F_{lift}\cdot\u\,dS.
\end{align*}

\subsection{Post-processing and data analysis}
The FEM model calculates tissue properties across a set of 128,000 quadrature points within each muscle block. We defined an orientation for the fibres at each quadrature point. The pennation angle $\beta_0$ in the undeformed and current $\beta$ states were calculated as the angle between the fibre orientations and the x-axis: this is an angle in 3D space, similar to the 3D pennation angles defined by Rana et al. 2013. We calculated forces $F$ as the magnitude of force perpendicular to a face on the muscle. The longitudinal muscle force in the line-of-action is denoted by $F_x$.

The strain-energies are initially calculated as strain energy-densities y, which are the strain- energy for a given volume of tissue, in units J m$^{-3}$ . We compute the total strain-energy of the tissue. The strain-energy potential $U$ is the strain-energy in the tissue, in units of Joules. We calculated $U$ at each given state by integrating y across the volume of muscle tissue at that state. We computed volumetric, muscle base-material, muscle active-fibre, and muscle passive-fibre strain-energy 
potentials: $U_{vol}$, $U_{base}$, $U_{act}$, $U_{pas}$, respectively (see Appendix A in Wakeling et al. 2020).

\section{Results}\label{section:results}
When the external load was applied to the passive muscle blocks on the z-face (‘top’ loading), the blocks decreased in their thickness in the z-direction, and also in their pennation angle. The extent of this compression increased with the external load. There was a minor effect of the pennation angle $\beta_0$ on this passive compression, with blocks at intermediate pennation angles of $\beta_0 = 20$ or 30 degrees
compressing slightly more than the parallel fibred block, or the $\beta_0 = 40$ degrees block. The increases in tissue
compression with increases in external transverse load are consistent with experimental measures using compression bandages (Wakeling et al. 2013) or weighted plungers on the medial gastrocnemius (Stutzig et al. 2019), and due to sitting on the gluteus maximus in humans (Linder-Ganz et al. 2007). Note that the internal pressure in the passive muscle block decreased with increasing external transverse load (Fig. 2b).

The internal pressure increased as the muscle activated. The internal pressure for the
maximum activation state of $\hat{a} = 1$ was 37 $\pm$ 20 kPa (mean $\pm$ S.D., $N=60$) across all geometries and transverse loads, which is within the range of intramuscular pressures measured for muscle contractions: 13-40 kPa in the frog gastrocnemius, 27 kPa for the soleus (Aratow et al. 1993) and 30 kPa for the tibialis anterior in humans (Ates et al. 2018). However, the external transverse load was not directly proportional to the increase in the internal pressure in the muscle blocks (Fig. 2b and 3). Indeed, the coefficient of determination between external transverse load and the internal pressure was $r^2 = 0.052$ for all states (N=600), and $r^2 = 0.042$ for the fully active state (N=60) in these
simulations.

The volumetric strain energy-density varies with both muscle length and pennation angles
(Fig. 4) and cannot be predicted just from the activation state of the muscle (Wakeling et al. 2020).
Across the range of pennation angles, activations, and muscle lengths used for the compression
simulations in this study, the coefficient of variation for the volume of the muscle blocks was 0.02.
By contrast, the range of the volumetric strain energy-densities of the muscle blocks was much larger
with a coefficient of variation of 1.00. It is this greater range of volumetric strain energy-densities
that drove the changes in internal pressure of the muscle (Fig. 3). Strain energy-density develops in
the active fibres within the muscle (that represent the contractile elements) when the activation state
increases, and is subsequently redistributed across the volumetric, base-material and passive-fibre
strain energy-densities (Wakeling et al. 2020). We have also found that the volumetric strain energy-
density varies with muscle length and pennation angle, and so too the internal pressure varies with
length and pennation angle. This is a result of the nearly incompressibility nature of our model. As
muscle tissues are allowed to change volume during contraction, the dilation $J$ changes, causing the
volumetric strain energy-density to change (see Equation 1 in Methods section for the formal
definition of this strain energy-density in terms of the dilation). Changes in both muscle length and
pennation cause changes in the volume and therefore local changes in the dilation. With the
definition of the internal pressure, we see that it also varies with muscle length and pennation angle.
However, we noted that internal pressure does not possess a direct relation with the external
transverse load: this is illustrated in Fig. 3 where each given transverse load can cause a range of
different internal pressures depending on the length or pennation angle of the muscle block being
compressed. Note that this highlights important considerations for interpreting experimental data
(Wakeling et al. 2013; Siebert et al. 2016, de Brito Fontana et al. 2018; Sleboda et al. 2019) where
muscle is compressed using transverse external loads, because the internal pressure in the muscle
may not be directly related to the extent of the external load.

The compressed muscle blocks changed in thickness when they were activated (Fig. 5A). The
least pennate muscle blocks ($\beta_0\leq 10$ degrees) increased in thickness (bulged) when they were activated, and
the more pennate blocks ($\beta_0\geq 20$ degrees) decreased in thickness. This difference in the direction of bulging
was consistent with previous experimental (Maganaris et al., 1998; Randhawa et al., 2013; Randhawa
and Wakeling, 2013, 2018; Raiteri et al. 2016) and modeling results (Wakeling et al. 2020). The
parallel fibred block of muscle ($\beta_0 = 10$ degrees) bulged during activation, however there was minimal effect of
the transverse load on this bulging. By contrast, contracting against greater transverse loads increased
the bulging and the pennation angle of the pennate muscle blocks (Fig. 5B). However, the pennation
angle never returned to its undeformed values of $\beta_0$ during these compressions and contractions. All
the pennate blocks ‘lifted’ the external load when it was applied from the ‘top’, or in other words, the
distance between the z-faces increased when the load was applied to the z-face. The ‘lifting work’
against this external load was the lowest for the $\beta_0 = 10$ degrees at 0.3 x 10 3 J m$^{-3}$, and increased to 1.4 x 10 3 J
m$^{-3}$ for $\beta_0 = 40$ degrees when measured in comparison to the unloaded state. This range of ‘lifting work’ spans the values recorded in experimental studies: 1.1-1.2 x 10 3 J m$^{-3}$ (Siebert et al. 2014) in rats, and 1.1 x10 3 J m$^{-3}$ in humans (Stutzig et al. 2019), where this work is expressed as a muscle volume-specific energy density.

The strain-energy increased in the muscle during contraction (Fig.4). The strain-energy
redistributed across different strain-energy potentials (volumetric, base-material, active-fibre and
passive-fibre) in a complex manner that depended on the muscle length, activation and pennation
angle, similar to our previous study (Wakeling et al. 2020). Muscles with greater initial pennation
angle $\beta_0$ developed much larger base-material strain-energy potentials (Fig. 4), due to the greater
shortening of the muscle fibres, in a manner also seen in our previous study (Wakeling et al. 2020).
The volumetric strain-energy potential was larger at longer muscle lengths $\hat{l}$ for $\beta_0\leq 20$ degrees as also
shown in our previous study (Wakeling et al. 2020), but the relation was more complex at $\beta_0\geq 30$ degrees
(Fig. 4). The volumetric strain-energy potential was reduced with greater transverse external loads for
most muscle lengths and pennation angles $\beta_0$, apart from at $\beta_0 = 40$ degrees and $\hat{l} = 1.0$, and at $\beta_0$ of 30-40 degrees
and $\hat{l} = 0.9$ (Fig. 4).

The force $F_x$ developed in the line-of-action of the muscle varied with muscle length,
pennation angle and external load (Fig. 6). For comparative purposes, the change in force $\Delta F_x$ is
expressed relative to the maximum uncompressed force for that muscle at length $\hat{l}$ and pennation
angle $\beta_0$ , and $\Delta F_x$ was normalized to the maximum uncompressed force that occurred at the resting
length $\hat{l} = 1.0$ for that pennation angle $\beta_0$. We also compared here the results for transverse loading
from the ‘top’ (z-) and the ‘side’ (y-) direction. The force $\Delta F_x$ was reduced with compression for
muscle blocks with $\beta_0\leq 20$ degrees, with the force reduction becoming greater for increased transverse
loads. The reduction in force was greater when the load was from the top than from the side. Patterns
of force reduction $\Delta F_x$ differed for the more pennate blocks: for the $\beta_0 = 40$ degrees block the force actually
increased with transverse loading from the top. For the transverse loading from the top, the muscle
blocks showed the greatest force reduction at their initial length $\hat{l} = 1.0$ for $\beta_0 = 0$ degrees. The greatest force
reduction at the short muscle length $\hat{l} = 0.9$ occurred at $\beta_0 = 10$ degrees, and the greatest force reduction at
the longest length $\hat{l} = 1.1$ occurred for $\beta_0 = 30$ degrees, with these patterns varying slightly when the load
was applied to the sides of the blocks. Here we show that the force reduction for muscle depends in a
complex manner on the length, pennation angle and direction of the transverse load, with this effect
being due to the way in which the strain-energy potentials are redistributed across the muscle during
these externally loaded contractions (Fig. 4).

The contributions of the different strain-energy potentials to the force $F_x$ in the line -of-action
of the muscle are shown in Fig. 7. For muscle with low to moderate initial pennation $\beta_0\leq 20$ degrees, the
largest reduction in force that occurred with external transverse load was from a reduced contribution
from the volumetric strain-energy potential. Reductions in the volumetric strain-energy potential
were less pronounced with external transverse load for the highest pennation angle of $\beta_0 = 40$ degrees,
however at this pennation angle the muscle force actually increased at the highest transverse load and
shortest length: this can be attributed to the substantial increase of the contribution from the base-
material strain-energy potential that occurred for this state.

\section{Discussion}
In this study we used the finite element method (FEM) to evaluate a 3D model of skeletal
muscle, based on the principles of continuum mechanics, to probe the relation between external
transverse load on the muscle and the force that it can develop in its line-of-action as well as the
internal work done by the tissues. The FEM model contained a series of constitutive relations that are
based on phenomenological descriptions of contractile elements and tissue properties (see details of
the model in Wakeling et al. 2020): none of these relations were specifically derived from or
optimized to the transverse response of muscle contractions, in contrast to previous models
(Randhawa and Wakeling, 2015; Siebert et al. 2012, 2014, 2018). Nonetheless, the model predicted
many of the general features of the compression response that have been previously reported, and so
these general features emerge from the physical principles that govern 3D deformations in muscle
tissue. We chose the main compression direction to be from the ‘top’ which is an external load acting
parallel to the plane of the muscle fibres and is the direction that was tested in previous uniaxial
loading in both animal (Siebert et al. 2014, 2016) and human experiments (Ryan et al. 2019; Stutzig
et al. 2019). With this compression from the top, our model predicted that passive muscle tissue
would decrease in thickness (Fig. 2), and the fibres would decrease in pennation angle, supporting
experimental results (Ryan et al. 2019). When the compressed muscle was activated, the model
predicted that it would increase in thickness to ‘lift’ the external load, generating lifting work (Siebert
et al. 2014) in the same range 0.3-1.4 x 10 3 J m$^{-3}$ as experimental measures 1.1-1.2 x 10 3 J m$^{-3}$
(Siebert et al. 2014; Stutzig et al. 2019), where this work is expressed as a muscle volume-specific
energy density.

The model in this study highlights the length-dependency of reductions in muscle force with
applied transverse loads. Most of our simulations showed a force reduction with transverse load, and
the extent of the force reduction was length-dependent at any given pennation angle. However, some
of the conditions at the highest pennation angles  $\beta_0 = 30-40$ degrees showed increases in force (Fig. 6). The
length dependency derives from both the fibre and the base properties of the muscle model. The
fibres are encoded as contractile elements that have length-dependent force properties for both the
active-fibre and passive-fibre components, and the base properties are governed by both the
volumetric and base-material relations (Rahemi et al. 2014; Wakeling et al. 2020). The combination
of the volumetric and base-material properties results in a tissue that tends to return to its initial state
(volume and shape) after it has been deformed, and this is a similar property to the helical-wound
representation of connective tissue modelled by Sleboda and colleagues (2019). It should be noted
that the initial undeformed state that these models return to is a discretionary choice between studies,
and so it should not be expected that the exact same length-dependency of the compression-force
reduction would occur across the different models. Indeed, this is the case where the helical model predicts force increase at longer lengths (Sleboda et al. 2019), whereas our model predicts these
increases at the largest pennation angles (Fig. 6).

The pennation angle of the muscle had a pronounced effect on the muscle response to
compression in terms of tissue deformation (Fig. 5), strain-energy potentials (Fig. 4), and the changes
in muscle force (Fig. 6). When pennate muscle contracts, the fibres rotate to greater pennation angles
as they shorten (Fukunaga et al. 1997; Maganaris et al. 1998). Muscle fibres act to draw the
aponeuroses together (or for these simulations, the z-faces) as they shorten, which tends to decrease
the muscle thickness. However, the fibres increase in girth during shortening in order to maintain
their volume (Rahemi et al., 2014, 2015). The increase in girth may be in either the width or
thickness direction, and indeed the relative deformations may vary between muscles (Randhawa and
Wakeling, 2015, 2018) due to stress asymmetries through the muscle (Wakeling et al. 2020).
However, a general effect is for the muscle to increase in pennation angle to allow their fibres to fit
within the enclosed volume of the muscle tissue (Zuurbier and Huijing, 1993; Fukunaga et al., 1997).
This increase in pennation angle tends to increase the muscle thickness (Randhawa and Wakeling,
2018), which in turn resists muscle compression acting from the ‘top’ direction and contributes to the
lifting work of the muscle. The results from these simulations support this explanation. We
additionally show how the strain-energy potentials redistribute within the muscle in a pennation-
dependent manner (Fig. 4; Wakeling et al. 2020). Thus, the response to the compression and the
internal work that can be done by the muscle will also be pennation dependent, due to the altered
balance of strain-energy potentials within the muscle. We show here that the force reduction that
occurs with transverse loading of the muscle seems particularly dependent on the volumetric strain-
energy potential (Fig. 7), that in turn varies with pennation angle and the direction of the external
load relative to the fibres (side or top: Fig. 4).

Strain-energy potentials develop during contraction and are distributed through the muscle
(Wakeling et al. 2020). When the muscle contracts it increases in its free energy, with this energy
being derived from the hydrolysis of ATP to ADP within the muscle fibres (Woledge et al., 1985;
Aidley, 1998). The active-fibre strain-energy potentials are redistributed to passive-fibre strain-
energy potentials and then to the base material strain-energy potential that develops in the bulk
muscle tissue within the muscle fibres (excluding the myofilament fraction), connective tissue
surrounding the muscle fibres such as the extracellular matrix, and in sheets of connective tissue that
form the aponeuroses and internal and external tendons. Energy is also used to change the muscle
volume. Whilst muscle is often assumed to be incompressible, small changes in volume can occur in
fibres (Neering et al., 1991), bundles of fibres called fascicles (Smith et al., 2011), and in whole
muscle (Bolsterlee et al., 2017): these changes in volume are energy-consuming processes. The
volumetric strain-energy potential, which accounts for an energetic penalty to any changes in volume
that occur, builds up as the muscle is activated and shows slight increases in volume (Wakeling et al.
2020). The transverse external loads in this study act to compress the volume of the muscle (Fig. 2A).
These changes in volume relate to changes to the volumetric strain-energy potential as the muscle is
compressed. The volumetric strain-energy potential contributes to the contractile force $F_x$ in the line-
of-action for all bar the $\beta_0=40$ degrees condition at its shortest length. Thus, the compression-induced
reductions in volumetric strain-energy potential result in the reductions to force in the line-of-action
during the muscle contractions (Fig. 7).

The volumetric strain-energy potential is arguably the least-well characterized component of
the internal energy in the muscle in our simulations. The extent of the increase in volume and the
volumetric strain-energy potential is related to the choice of the bulk modulus $\kappa$ of the tissue. A
constitutive equation to calculate the volumetric strain-energy potential has not been defined for muscle tissue, and so we used a general form (equation 1) that is used for compressible neo-Hookean
material (see, e.g. Pelteret 2012). Here we used a value of $\kappa = 10^6$ Pa that was consistent with
previous studies (Rahemi et al., 2014, 2015, Wakeling et al. 2020). We previously showed that this k
resulted in volume changes of 2\%-4\% during contraction of fully active parallel muscle fibres.
Nonetheless, a previous study showed that k can be varied across a wide range of magnitudes and
still result in similar predictions of tissue deformation (Gardiner and Weiss, 2001). Given the
apparent importance of the volumetric strain-energy potential to the modulation of contractile force
in response to muscle compression, establishing muscle-specific constitutive equations for the
volumetric strain energy-density and values for the bulk modulus will be an important area of future
investigation.


%

\paragraph{Conflict of Interest} The authors declare that the research was conducted in the absence of any commercial or financial
relationships that could be construed as a potential conflict of interest.

\paragraph{Author Contributions} DR, NN, JW contributed to the study design. DR, SD, SR, NN, JW contributed to the model
development. DR ran all the simulations for the paper and data analysis. DR and JW contributed to
the first draft of the manuscript. DR, SD, SR, NN, JW contributed to final manuscript preparation.

\paragraph{Funding} We thank the Natural Sciences and Engineering Research Council of Canada for Discovery Grants to J.M.W. and N.N., and an Alexander Graham Bell Canada Graduate Scholarship-Doctoral to S.A.R. We are also grateful for funding to S.D. from Comisión Nacional de Investigación Científica y Tecnológica of Chile through Becas-Chile.

\paragraph{Acknowledgments}
We thank Tobias Siebert for extensive discussions and inspiration about the mechanics of muscle compression.

\section*{References}
\hspace{15pt} Aidley, D. J. (1998). The Physiology of Excitable Cells, 4th Edition. Cambridge: Cambridge
University Press.

Aratow, M., Ballard, R., Crenshaw, A., Styf, J., Watenpaugh, D., Kahan, N., and Hargens, A. (1993).
Intramuscular pressure and electromyography as indexes of force during isokinetic exercise.
J. Appl. Physiol., 74, 2634-2640.

Azizi, E., Brainerd, E., and Roberts, T. (2008). Variable gearing in pennate muscles. PNAS, 105(5),
1745-1750.

Azizi, E., Deslauriers, A. R., Holt, N. C., and Eaton, C. E. (2017). Resistance to radial expansion
limits muscle strain and work. Biomech. Model. Mechanobiol.16, 1633-1643. doi:
10.1007/s10237-017-0909-3

Böl, M., Leichsenring, K., Weichert, C., Sturmat, M., Schenk, P., Blickhan, R. and Siebert, T.
(2013). Three-dimensional surface geometries of the rabbit soleus muscle during contraction:
input for biomechanical modelling and its validation. Biomech. Model. Mechanobiol. 12,
1205-1220. doi: 10.1007/s10237-013-0476-1

Bolsterlee, B., D'Souza, A., Gandevia, S. C., and Herbert, R. D. (2017). How does passive
lengthening change the architecture of the human medial gastrocnemius muscle? J. Appl.
Physiol. 122, 727-738. doi: 10.1152/japplphysiol.00976.2016

Brainerd, E. L., and Azizi, E. (2005). Muscle fiber angle, segment bulging and architectural gear ratio
in segmented musculature. J. Exp. Biol. 208, 3249–3261. doi: 10.1242/jeb.01770

de Brito Fontana, H., de Campos, D., Sawatsky, A., Han, S.-W., and Herzog, W. (2020). Why do
muscles lose torque potential when activated within their agonistic group? J. Exp. Biol.,
223(1), jeb213843. doi:10.1242/jeb.213843

de Brito Fontana, H., Han, S. W., Sawatsky, A., and Herzog, W. (2018). The mechanics of agonistic
muscles. J. Biomech. 79, 15-20. doi: 10.1016/j.jbiomech.2018.07.007

Dick, T. J. M., and Wakeling, J. M. (2017). Shifting gears: dynamic muscle shape changes and force-
velocity behavior in the medial gastrocnemius. J. Appl. Physiol. 123, 1433–1442. doi:
10.1152/japplphysiol.01050.2016

Fukunaga, T., Ichinose, Y., Ito, M., Kawakami, Y., and Fukashiro, S. (1997). Determination of
fascicle length and pennation in a contracting human muscle in vivo. J. Appl. Physiol., 82(1),
354–358. doi:10.1152/jappl.1997.82.1.354

Gardiner, J. C., and Weiss, J. A. (2001). Simple Shear Testing of Parallel-Fibered Planar Soft
Tissues. J. Biomech. Eng. 123, 170–175. doi: 10.1115/1.1351891

Linder-Ganz, E., Shabshin, N., Itzchak, Y., Gefen, A., (2007). Assessment of mechanical conditions
in sub-dermal tissues during sitting: a combined experimental-MRI and finite element
approach. J. Biomech. 40, 1443–1454.

Maganaris, C., Baltzopoulos, V., and Sargeant, A. (1998). In vivo measurements of the triceps surae
complex architecture in man: implications for muscle function. J. Physiol. 512, 603-614. doi:
10.1111/j.1469-7793.1998.603be.x

Neering, I. R., Quesenberry, L.A., Morris, V.A., and Taylor, S.R. (1991). Nonuniform volume
changes during muscle contraction. Biophys. J. 59, 926–933. doi: 10.1016/S0006-
3495(91)82306-2

Pelteret, J. P., and McBride, A. (2012). The deal.II tutorial step-44: Three-field formulation for non-
linear solid mechanics. doi: 10.5281/zenodo.439772

Rahemi, H., Nigam, N., and Wakeling, J. M. (2014). Regionalizing muscle activity causes changes to
the magnitude and direction of the force from whole muscles – a modeling study. Front.
Physiol. 5:298. doi: 10.3389/fphys.2014.00298

Rahemi, H., Nigam, N., and Wakeling, J. M. (2015). The effect of intramuscular fat on skeletal
muscle mechanics: implications for the elderly and obese. J. R. Soc. Interface 12:20150364.
doi: 1098/rsif.2015.0365

Raiteri B.J., Cresswell A.G., and Lichtwark G.A. (2016) Three-dimensional geometrical changes of
the human tibialis anterior muscle and its central aponeurosis measured with three-
dimensional ultrasound during isometric contractions. PeerJ 4:e2260.

Raiteri, B. J., Cresswell, A. G. and Lichtwark, G. A. (2016). Three-dimensional geometrical changes
of the human tibialis anterior muscle and its central aponeurosis measured with three-
dimensional ultrasound during isometric contractions. PeerJ. 4:e2260. doi:
10.7717/peerj.2260

Rana, M., Hamarneh, G., and Wakeling, J. (2013). 3D fascicle orientations in triceps surae. J. Appl
.Physiol. 115: 116–125, 2013

Randhawa, A., and Wakeling, J. M. (2015). Multidimensional models for predicting muscle structure
and fascicle pennation. J. Theor. Biol. 382, 57–63. doi: 10.1016/j.jtbi.2015.06.001

Randhawa, A., and Wakeling, J. M. (2018). Transverse anisotropy in the deformation of the muscle
during dynamic contractions. J. Exp. Biol. 221:jeb175794. doi: 10.1242/jeb.175794

Randhawa, A., Jackman, M. E., and Wakeling, J. M. (2013). Muscle gearing during isotonic and
isokinetic movements in the ankle plantarflexors. Eur. J. Appl. Physiol. 113, 437–447. doi:
10.1007/s00421-012-2448-z

Ryan, D. S., Stutzig, N., Siebert, T., \& Wakeling, J. M. (2019). Passive and dynamic muscle
architecture during transverse loading for gastrocnemius medialis in man. J. Biomech. 86,
160–166. doi:10.1016/j.jbiomech.2019.01.054

Siebert, T., Till, O., \& Blickhan, R. (2014). Work partitioning of transversally loaded muscle:
Experimentation and simulation. Comp. Meth. Biomech. Biomedi. Eng. DOI:
10.1080/10255842.2012.675056

Siebert, T., Günther, M., \& Blickhan, R. (2012). A 3D-geometric model for the deformation of a
transversally loaded muscle. J. Theor. Biol. 298(C), 116–121. doi:10.1016/j.jtbi.2012.01.009

Siebert, T., Stutzig, N., and Rode, C. (2018). A hill-type muscle model expansion accounting for
effects of varying transverse muscle load. J. Biomech. 66, 57–62. \\
doi:10.1016/j.jbiomech.2017.10.043

Sleboda, D. A., and Roberts, T. J. (2019). Internal fluid pressure influences muscle contractile force.
PNAS, 5(3), 201914433–1778. doi:10.1073/pnas.1914433117

Sleboda, D. A., and Roberts, T. J. (2017). Incompressible fluid plays a mechanical role in the
development of passive muscle tension. Biol. Lett. 13, 20160630. doi:
10.1098/rsbl.2016.0630

Smith, L. R., Gerace-Fowler, L., and Lieber, R. L. (2011). Muscle extracellular matrix applies a
transverse stress on fibers with axial strain. J. Biomech. 44, 1618–1620. doi:
10.1016/j.jbiomech.2011.03.009

Stutzig, N., Ryan, D., Wakeling, J. M., and Siebert, T. (2019). Impact of transversal calf muscle
loading on plantarflexion. J. Biomech. 85, 37–42. doi:10.1016/j.jbiomech.2019.01.011

Wainwright S. A., Biggs W. D., Currey J. D., and Gosline J. M. (1976). Mechanical design in
organisms. John Wiley \& Sons Inc

Wakeling J. M., Ross S. A., Ryan D. S., Bolsterlee B., Konno R., Domínguez S., and Nigam N.
(2020). The energy of muscle contraction. I. Tissue force and deformation during isometric
contractions. Submitted to Frontiers Physiology (special project on muscle ECM).

Wakeling, J. M., Jackman, M., and Namburete, A. I. (2013). The effect of external compression on
the mechanics of muscle contraction. J. Appl. Biomech. 29, 360–364. doi:
10.1123/jab.29.3.360

Weis-Fogh, T., and Alexander, R. M. (1977). “The sustained power output from straited muscle,” in
Scale Effects in Animal Locomotion. ed T. J. Pedley (New York: Academic Press), 511-525.

Woledge, R. C., Curtin, N. and Homsher, E. (1985). Energetic Aspects of Muscle Contraction.
Monogr. Physiol. Soc. 41, 1-357.

Zuurbier, C. J., and Huijing, P. A. (1993). Changes in muscle geometry of actively shortening
unipennate rat gastrocnemius muscle. J. Morphol. 218, 167–180.\\
doi:10.1002/jmor.1052180206

\clearpage
\section*{Figures}
\graphicspath{{./Figs/}}
\begin{figure}[!ht]
    \centering
    \includegraphics[width = 1\textwidth, 
height=.8\textheight]{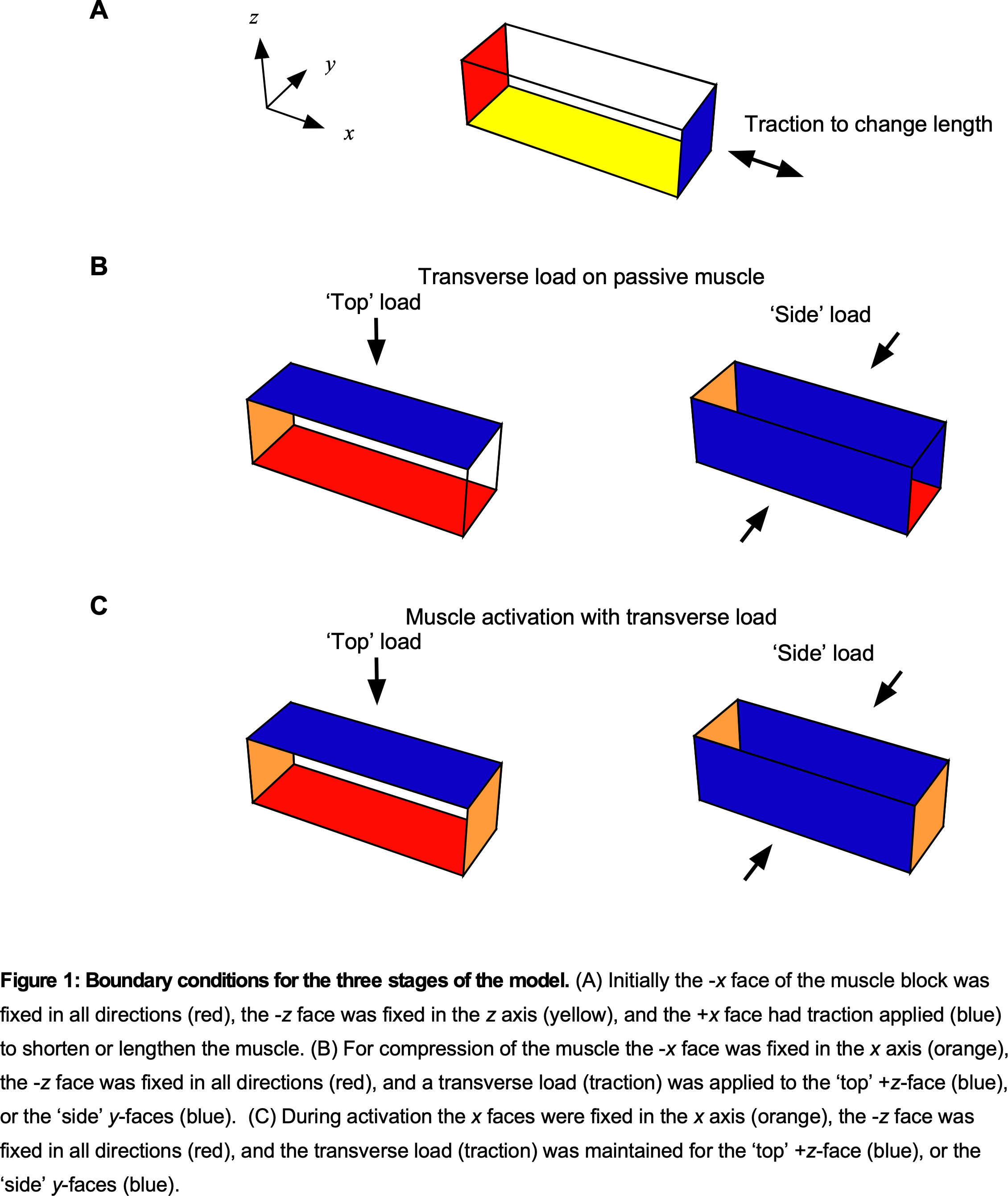}
\end{figure}
\begin{figure}[!ht]
    \centering
    \includegraphics[width = 1\textwidth, 
height=.8\textheight]{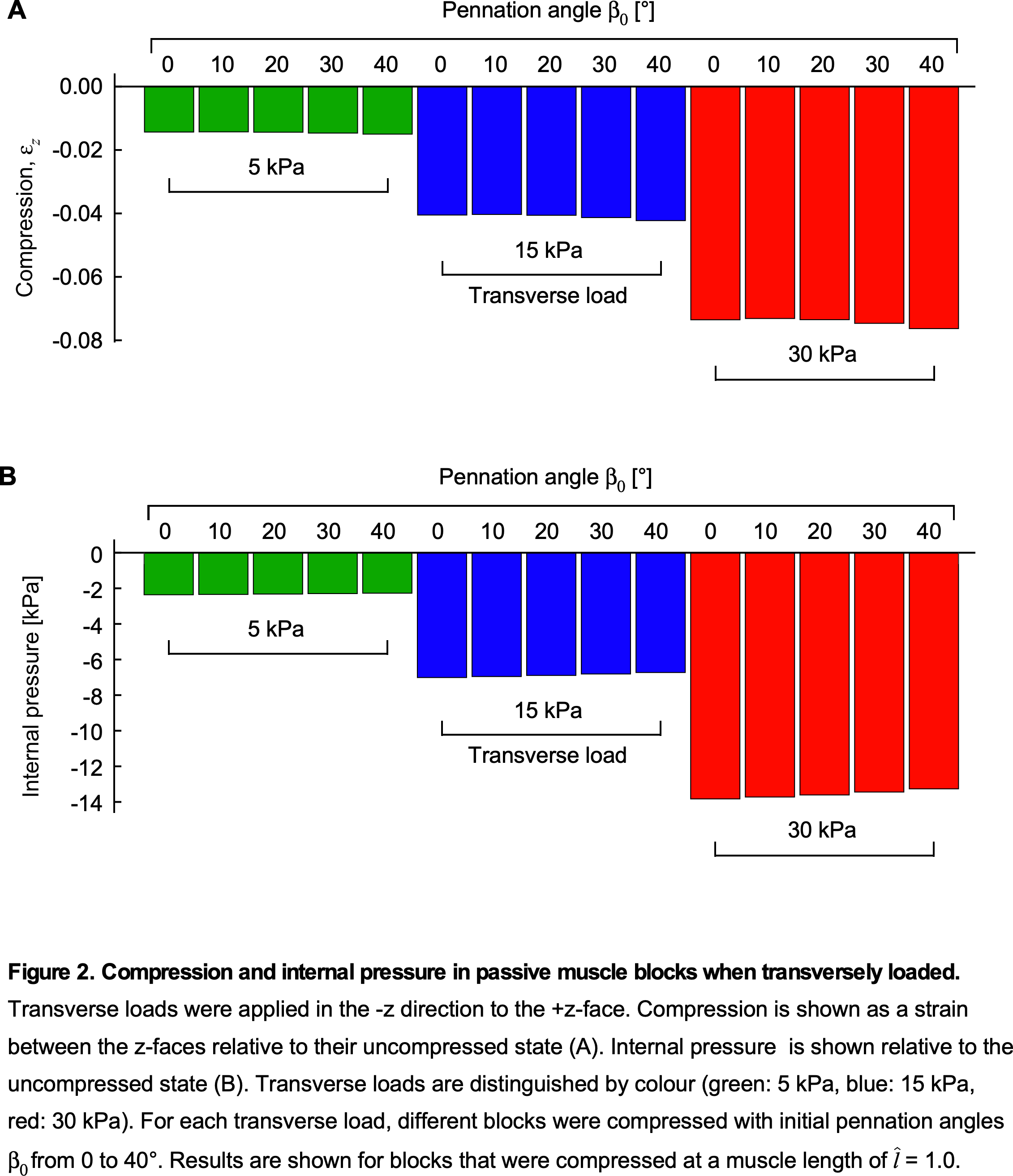}
\end{figure}
\begin{figure}[!ht]
    \centering
    \includegraphics[width = 1\textwidth, 
height=.8\textheight]{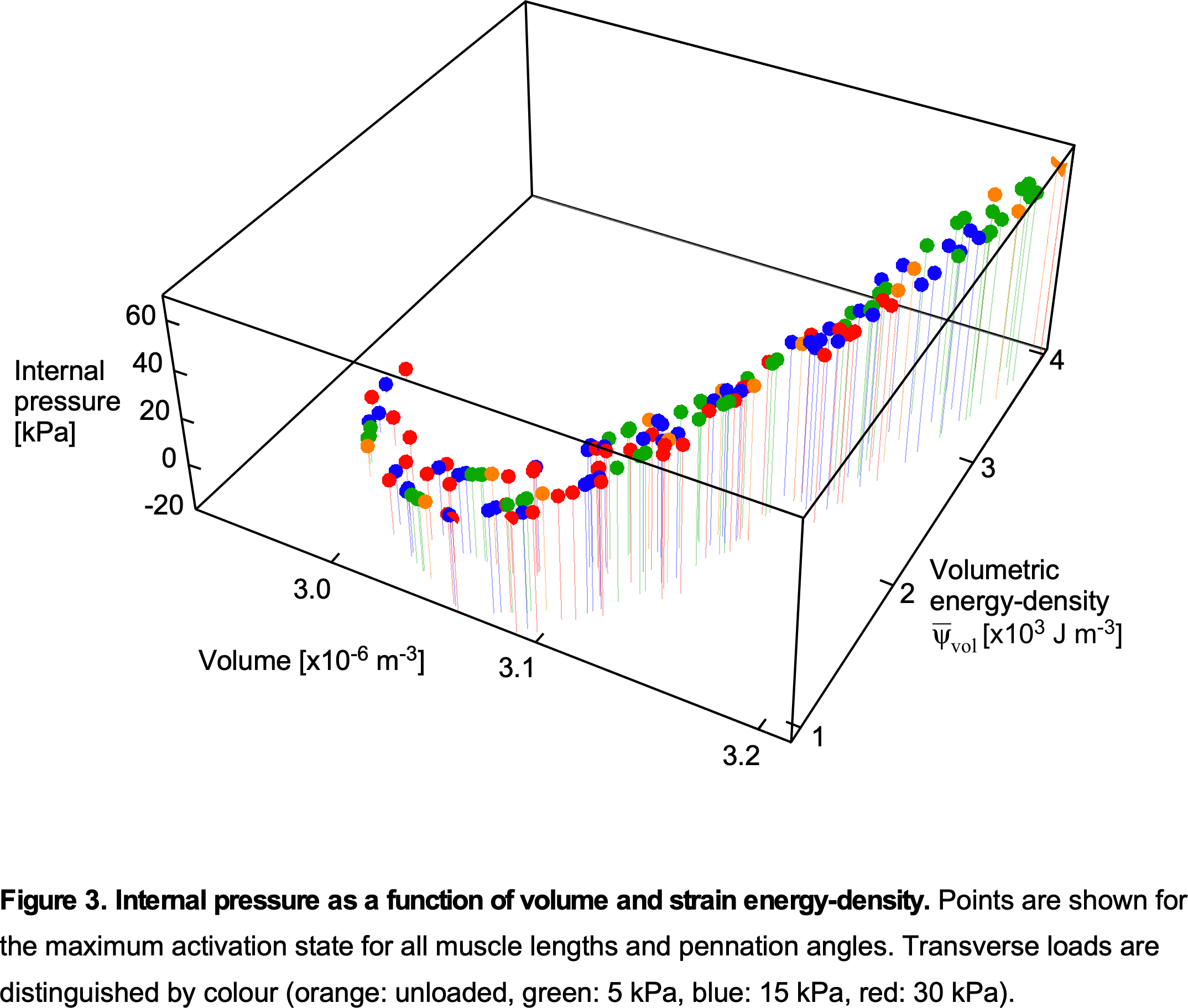}
\end{figure}
\begin{figure}[!ht]
    \centering
    \includegraphics[width = 1\textwidth, 
height=1\textheight]{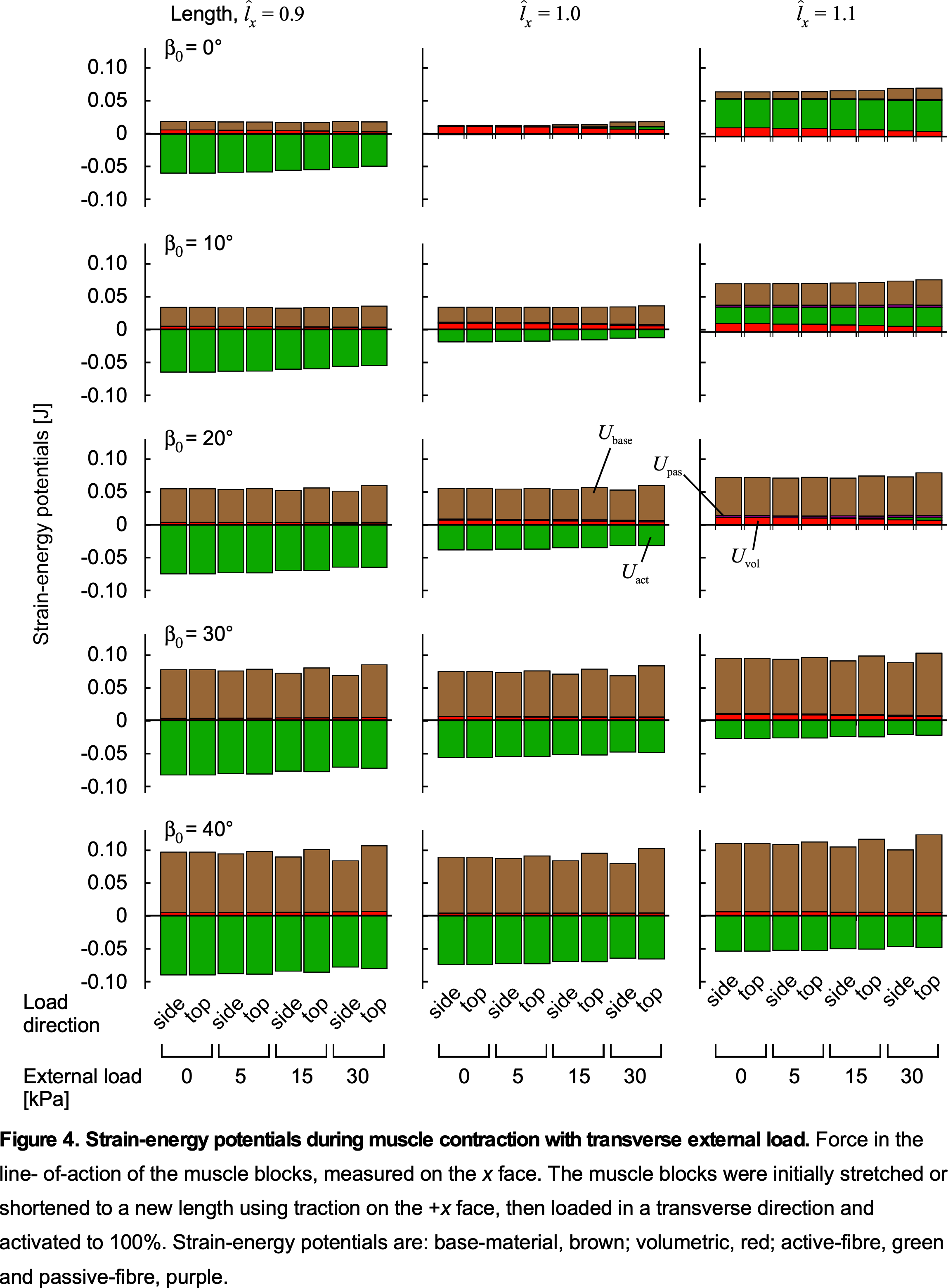}
\end{figure}
\begin{figure}[!ht]
    \centering
    \includegraphics[width = 1\textwidth, 
height=1\textheight]{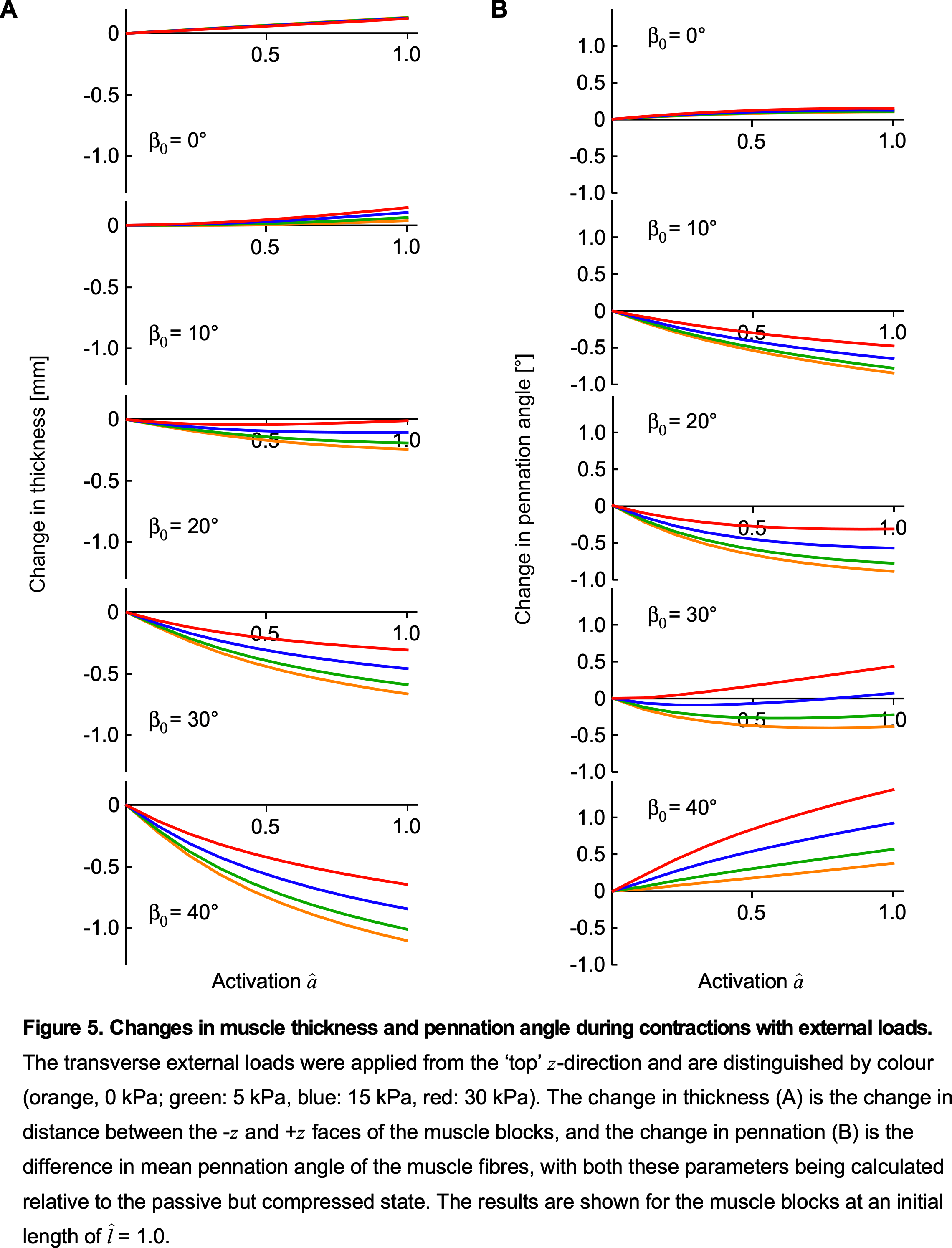}
\end{figure}
\begin{figure}[!ht]
    \centering
    \includegraphics[width = 1\textwidth, 
height=1\textheight]{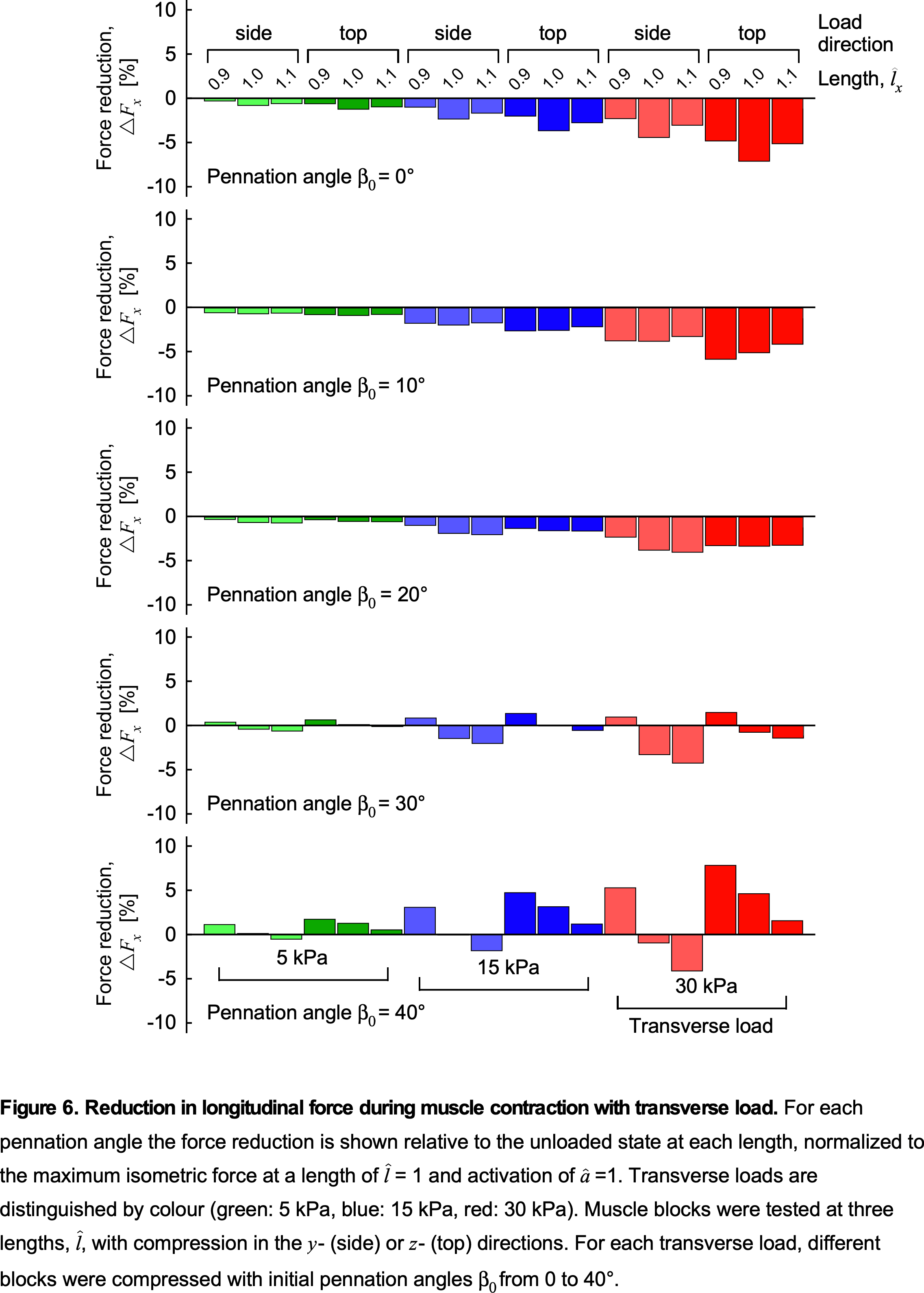}
\end{figure}
\begin{figure}[!ht]
    \centering
    \includegraphics[width = 1\textwidth, 
height=1\textheight]{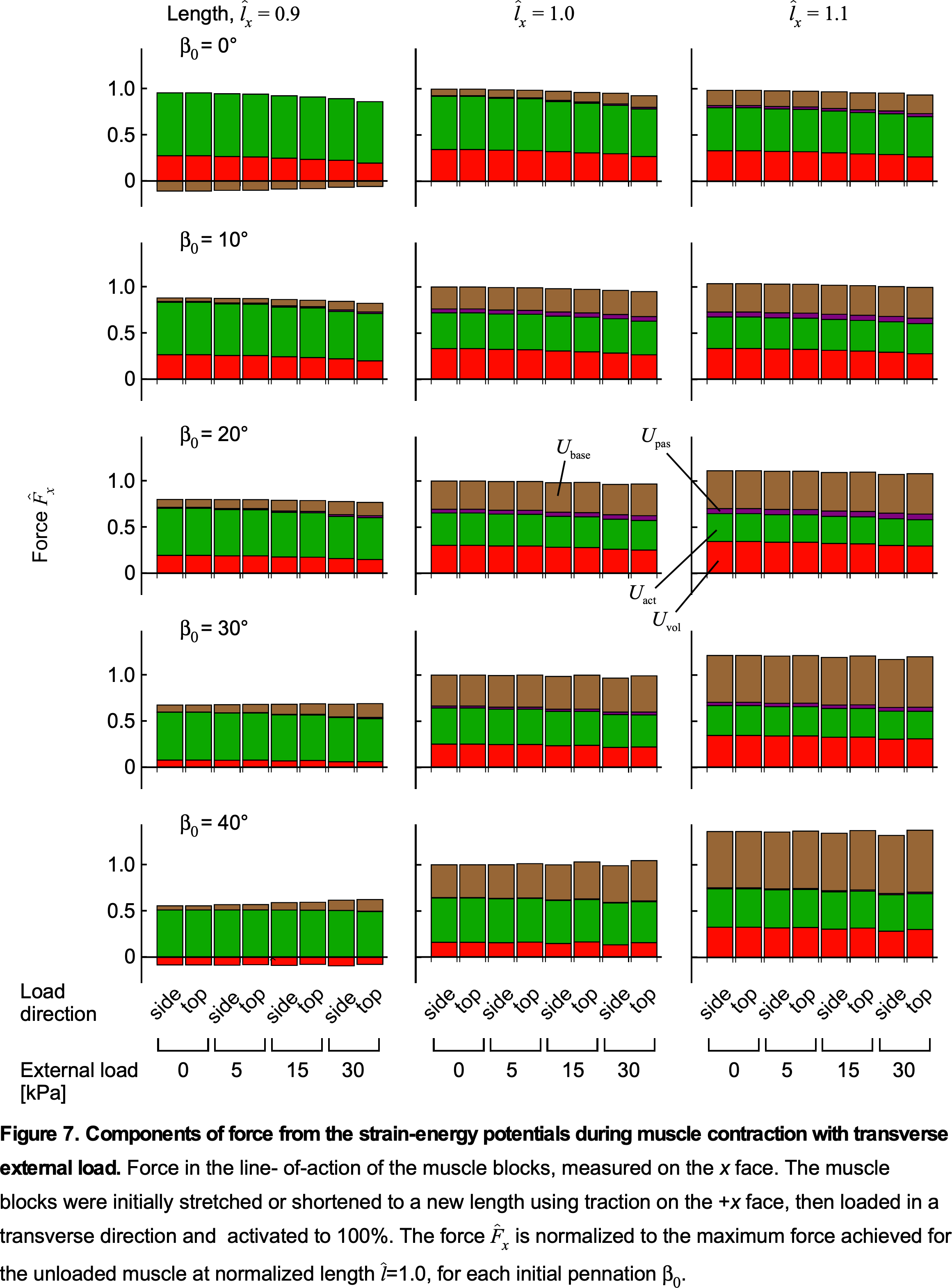}
\end{figure}
\end{document}